\documentclass[12pt]{spieman}  
\usepackage{amsmath,amsfonts,amssymb}
\usepackage{setspace}
\usepackage{tocloft}
\usepackage{multicol,tcolorbox,hhline}
\usepackage{graphicx}

\newcommand{\ee}[1]{\mbox{${} \times 10^{#1}$}}
\newcommand{\eten}[1]{\mbox{$10^{#1}$}}
\newcommand{\msun}{\mbox{M$_\odot$}}
\newcommand{\mjup}{\mbox{M$_J$}}
\newcommand{\um}{$\mathrm{\mu}$m}

\def\kms{{km s$^{-1}$}}
\def\micron{{$\mu$m}}

\def\SALT{\emph{SALTUS}}

\newcommand{\water}{H$_2$O}

\newcommand{\cotwo}{CO$_2$}

\newcommand{\hh}{\mbox{{\rm H}$_2$}}
\newcommand{\nn}{\mbox{{\rm N}$_2$}}

\newcommand{\ammonia}{\mbox{{\rm NH}$_3$}}

\newcommand{\methanol}{\mbox{\rm CH}$_3$\mbox{\rm OH}}
\newcommand{\methane}{\mbox{\rm CH}$_4$}

\title{Star and Planet Formation with the Single Aperture Large Telescope for Universe Studies (SALTUS) Space Observatory}

\author[a]{Kamber R. Schwarz}
\author[b]{Alexander Tielens}
\author[c]{Joan Najita}
\author[d]{Jennifer Bergner}
\author[e]{Quentin Kral}
\author[f]{Carrie Anderson}
\author[f]{Gordon Chin}
\author[g]{David Leisawitz}
\author[h]{David Wilner}
\author[i]{Peter Roelfsema}
\author[i]{Floris van der Tak}
\author[j]{Erick Young}
\author[k]{Chris Walker}
\affil[a]{Max-Planck-Institut f\"{u}r Astronomie (MPIA), K\"{o}nigstuhl 17, 69117 Heidelberg, Germany}
\affil[b]{Astronomy Department, University of Maryland, 296 Stadium Dr. PSC, College Park, USA, 20742}
\affil[c]{National Optical Astronomy Observatory (NOAO), 950 N. Cherry Ave, Tucson, AZ 85719, USA}
\affil[d]{University of California, Berkeley, Berkeley, CA, 94720, USA}
\affil[e]{LESIA, Observatoire de Paris, Université PSL, CNRS, Sorbonne Université, Université Paris Cité, 5 place Jules Janssen,
92195 Meudon, France}
\affil[f]{NASA Goddard Space Flight Center, Planetary Systems Laboratory, 8800 Greenbelt Road, Greenbelt, USA, 20771}
\affil[g]{NASA Goddard Space Flight Center, Observational Cosmology Laboratory, 8800 Greenbelt Road, Greenbelt, MD USA 20771}
\affil[h]{Center for Astrophysics Harvard Smithsonian, 60 Garden St., Cambridge, MA 92138, USA}
\affil[i]{Netherlands Institute for Space Research (SRON), Landleven 12, 9747 AD Groningen, The Netherlands}
\affil[j]{Universities Space Research Association, 425 3rd Street SW, Suite 950, Washington DC, 20024, USA}
\affil[k]{University of Arizona, Department of Astronomy and Steward Observatory, Tucson, USA, 85719}

\cftpagenumbersoff{figure}
\cftpagenumbersoff{table} 
\begin{document} 
\maketitle

\begin{abstract}
 The Single Aperture Large Telescope for Universe Studies (\SALT) is a far-infrared space mission concept with unprecedented spatial and spectral resolution. \SALT\ consists of a 14-m inflatable primary, providing $16\times$ the sensitivity and $4\times$ the angular resolution of \emph{Herschel}, and two cryogenic detectors spanning a wavelength range of 34-660 \um\ and spectral resolving power of $300 - 10^{7}$. 
 Spectroscopic observations in the far-infrared offer many unique windows into the processes of star and planet formation. These include observations of low energy water transitions, the \hh\ mass tracer HD, many CHONS constraining molecules such as \ammonia\ and H$_2$S, and emission lines from the phonon modes of molecular ices. Observing these species will allow us to build a statistical sample of protoplanetary disk masses, characterize the water snowline, identify Kuiper Belt like debris rings around other stars, and trace the evolution CHONS from prestellar cores, through to protoplanetary disks and debris disks. This paper details details several key star and planet formation science goals achievable with \SALT.

\end{abstract}


{\noindent \footnotesize\textbf{*}Kamber R. Schwarz,  \linkable{schwarz@mpia.de} }

\begin{spacing}{2}   

\section{Introduction}
\label{sect:intro}  

The Single Aperture Large Telescope for Universe Studies (\SALT) is a far-infrared space mission concept proposed to NASA under the Astrophysics Probe Explorer (APEX) Announcement of Opportunity in November 2023. 
\SALT\ covers the far-infrared wavelength range $\approx 30-700$~\um, most of which is not covered by any current observatory. The design of \SALT\ consists of a 14m off-axis inflatable primary aperture and two cryogenic instruments: SAFARI-Lite and the High Resolution Receiver (HiRX). The large aperture size allows for unprecedented sensitivity and a spatial resolution of $\sim 1$\ensuremath{''} at 50~$\mu$m". 
The full technical details of the \SALT\ observatory can be found in Arenberg et al., ``Design, Implementation and Performance of the Primary Reflector for SALTUS," Kim et al, ``SALTUS Observatory Optical Design and Performance," and Harding, Arenberg, Donovan et al.,“SALTUS Probe Class Space Mission: Observatory Architecture \& Mission Design,” \textit{J. Astron. Telesc. Instrum. Syst.} (this issue).
SAFARI-Lite is a direct-detection grating spectrometer providing simultaneous 35–230~\um\ spectroscopy with a resolving power of $R=300$. The full technical details can be found in Roelfsema et al., “The SAFARI-Lite Imaging Spectrometer for the \SALT\ Space Observatory,” \textit{J. Astron. Telesc. Instrum. Syst.} (this issue).
HiRX is a multi-pixel, multi-band heterodyne receiver system spanning wavelength ranges 522--659 $\mu$m, 136--273 $\mu$m, 111.9--112.4 $\mu$m, 63.1--63.4 $\mu$m, and 56.1--56.4 $\mu$m with a resolving power of $R= 1\times 10^5 - 1\times 10^7$. The full techincal details can be found in Walker et al., “The High Resolution Receiver (HiRX) for the Single Aperture Large Telescope for Universe Studies (SALTUS),” \textit{J. Astron. Telesc. Instrum. Syst.} (this issue).

This paper provides an overview of the promise of \SALT\ for understanding star and planet formation, including molecular clouds, protostellar cores, protoplanetary disks, and debris disks. Accompanying papers in this issue describe the plans for guaranteed-time (GTO) and guest observing (Chin et al., “Single Aperture Large Telescope for Universe Studies (SALTUS): Probe Mission and Science Overview,” \textit{J. Astron. Telesc. Instrum. Syst.}), SALTUS’ contributions to High-Redshift Science (Spilker et al., ``Distant Galaxy Observations''\textit{J. Astron. Telesc. Instrum. Syst.}), Milky Way and nearby galaxies science (Levy et al., ``Nearby Galaxy Observations'' \textit{J. Astron. Telesc. Instrum. Syst.}), and solar system observations (Anderson et al., ``Solar System Science'' \textit{J. Astron. Telesc. Instrum. Syst.}). Additionally, some of \emph{SALTUS's} key science cases build on the \emph{OASIS} MIDEX-class mission concept, which used a similar large inflatable aperture for terahertz frequency observations \cite{Walker2021}. 

\subsection{Programmatic Motivation}
\label{sect:motivation} 

The \SALT\ star and planet formation science programs presented here address multiple high-priority science questions as identified by Astro2020 \cite{roadmap}, detailed below.
\begin{itemize}
\item Question E-Q1c: \emph{How Common Is Planetary Migration, How Does It Affect the Rest of the Planetary System, and What Are the Observable Signatures?} \SALT\ will provide the disk water measurements, including measurements of the water snowline location,
needed to connect atmosphere compositions to the water distribution of planet-forming disks and thereby to connect JWST observations of exoplanet atmospheres to a formation time and location. (\S\ref{sec:diskwater})

\item Question E-Q1d: \emph{How Does the Distribution of Dust and Small Bodies in Mature Systems Connect to the Current and Past Dynamical States Within Planetary Systems?} \SALT\ SAFARI-Lite will determine the occurance of exo-Kuiper belts around the nearest 30 G and K stars known to host debris disks, characterizing the commonality of dust in mature planet systems (\S\ref{sec:exokuiperbelt}).
\item Question E-Q3a: \emph{How Are Potentially Habitable Environments Formed?} \SALT\ will answer this question by observing [CII] at 157 \micron\ and [OI] at 63 and 145 \micron\ in debris disks, tracing the C/O ratio of material available for accretion onto terrestrial planets (\S\ref{sec:debrisdiskCO}).
\item Question E-Q3b: \emph{What Processes Influence the Habitability of Environments?} and Question F-Q4b: \emph{What Is the Range of Physical Environments Available for Planet Formation?} \SALT\ will determine the mass and temperature structure, as well as the abundance of CHONS bearing species, in roughly 1000 protoplanetary disks across evolutionary stages.   (\S\ref{sec:HDdiskmass},\S\ref{sec:diskwater},\S\ref{sec:diskastrochem})
\end{itemize}

\section{Star and Planet Formation Science with SALTUS}\label{sec:science} 

\subsection{Protoplanetary Disk Mass}\label{sec:diskmass} 
One of the most fundamental properties of planet formation is the mass of a planet-forming disk, which determines the total amount of material available to forming planets and the mechanisms through which planets can form, e.g, through gravitational instability vs. via core accretion \cite{Armitage2020}. The main contributor to the disk mass is \hh, which does not emit for the majority of disk regions since the molecule has no permanent dipole moment, with large energy spacings not well matched to the local temperatures. The ground-state transition is the quadrupole J=2-0 with an energy spacing of 510 K. Thus, exciting an \hh\ molecule to the J=2 state requires high gas temperatures and \hh\ emission originates only from the illuminated surface layers of the disk within a fraction of an au of the central star. Since most of the gas is at larger radii and is much colder, alternate tracers must be used to determine the total gas mass.

The most commonly used gas mass tracers in protoplanetary disks are continuum emission from dust and emission from rotational transitions of CO. Each method relies on different problematic assumptions. Uncertainties in the dust grain optical properties and the grain size distribution lead to significant uncertainty in the derived dust mass from observed emission. Then, to convert from dust mass to gas mass, a gas-to-dust mass ratio must be assumed. This value is typically assumed to be 100, as has been measured in the interstellar medium (ISM). However, several factors can change this ratio in disks, including loss of gas due to disk winds and accretion onto the central star, which will decrease the gas-to-dust ratio, and growth of dust grains beyond cm sizes, at which point the dust emission is no longer observable. Additionally, assuming a constant gas-to-dust ratio across the disk is not appropriate since high spatial resolution observations at millimeter wavelengths demonstrate that the outer radius of the dust disk is often much smaller than the outer radius of the gas disk \cite{Bergin2013,McClure2016}.

The CO abundance relative to \hh\ in the ISM is well constrained to be 5\ee{-5} to 2\ee{-4} \cite{Du2015}. However, when converting from CO abundance to \hh\ in a protoplanetary disk, additional corrections must be made to account for the reduced abundance of CO relative to \hh\ in the surface layer, where CO is photo-dissociated, and near the cold midplane, where CO is frozen out onto dust grains \cite{Miotello14}. Additional chemical reactions in the gas and on dust grains can also destroy CO \cite{Schwarz2018}. The resulting reduction in CO gas abundance, whatever the cause, varies not only across sources but also as a function of radius within a single disk \cite{Zhang2019}. Thus, there are large uncertainties when converting CO flux to total gas mass \cite{Miotello2023}.

Given the myriad assumptions that go into each technique, it is not surprising that the two methods of determining disk mass rarely agree. Alternative mass probes, preferably requiring fewer assumptions, are needed to determine the true disk gas mass. One possibility is to use the disk rotation curve to constrain the enclosed mass \cite{Veronesi2021}. However, because disks must always be less massive than the central star in order to remain gravitationally stable, the contribution of the disk to the rotation curve is small. This technique is only feasible for a small number of the most massive disks \cite{Andrews2024}.

\subsubsection{Tracing Mass with HD}\label{sec:HDdiskmass} 
\SALT\ will use HD to measure the gas mass
in hundreds of disks, establishing the variation
in this fundamental parameter across systems.
Observations of the \hh\ isotopologue HD
are unique to the far-IR, with the ground state
1-0 rotation transition at 112.07~\micron\ (2.675 THz;
SAFARI-Lite LW Band, HiRX-Band 3) 
and the
2-1 transition at 56.24~\micron\ (5.331 THz; SAFARI-
Lite MW Band, HiRX-Band 4b). HD is
the main reservoir of deuterium and its abundance
relative to \hh\ will be close to the elemental
D/H abundance; thus, HD can be used
to trace disk mass while avoiding many of the
limitations of other mass tracers. For example,
HD emission is optically thin and not subject to
chemical processing that can change abundances
of other tracers relative to \hh\ \cite{Bergin17}.
Observing of only the HD 1-0 are capable of constraining the disk mass to within a factor of 2-10 depending on disk mass, while the additional observation of the HD 2-1 line decreases this uncertainty to no more than a factor of 3 \cite{Trap2017,Kamp2021}. \SALT\ HiRX is designed to observe both lines simultaneously. 

\begin{figure}
\begin{center}
\begin{tabular}{c}
\hspace{-3mm}\includegraphics[height=5.0cm]{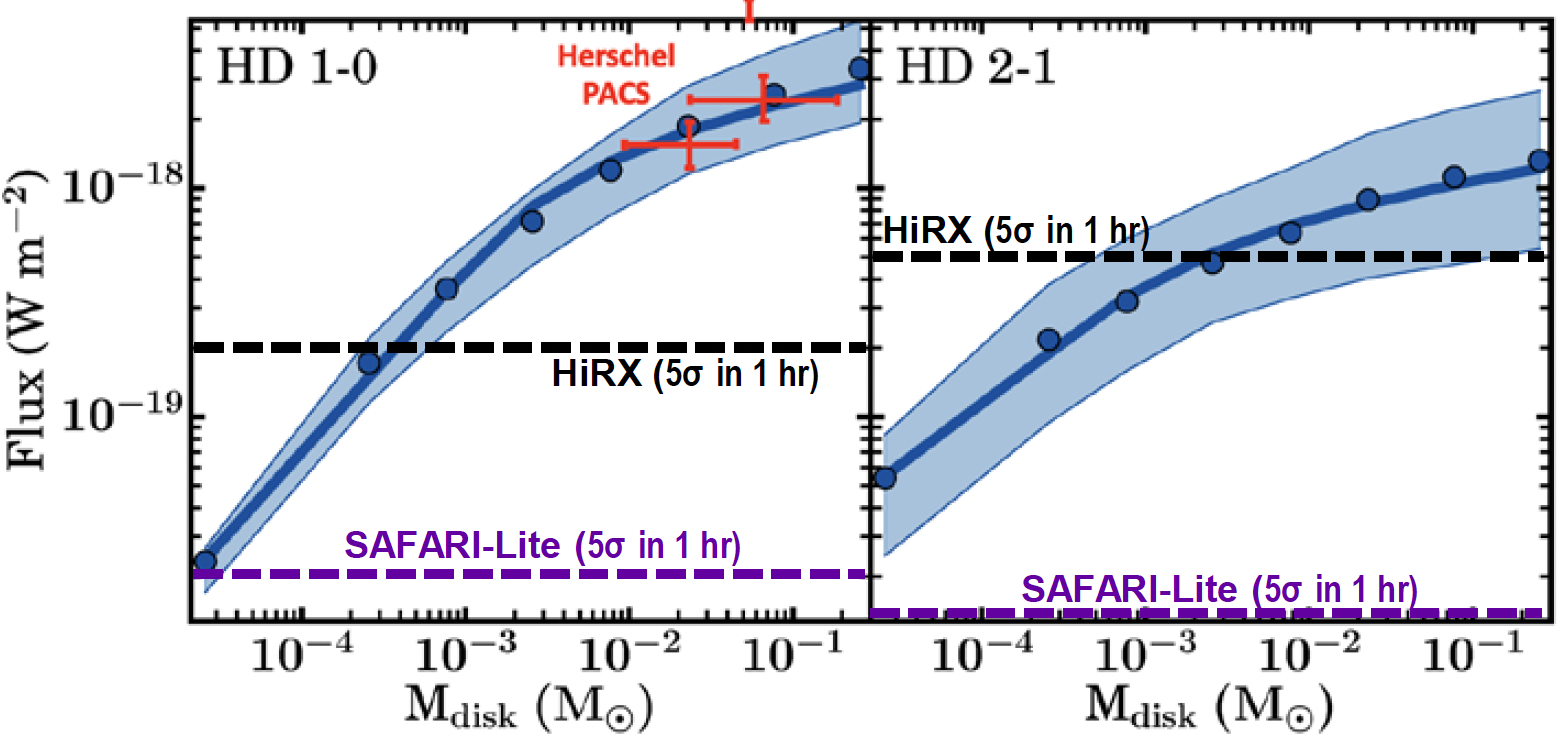} 
\end{tabular}
\end{center}
\caption 
{\label{fig:trapman}Model predictions of disk gas (D=140 pc), as a function of HD line flux. The width of the blue band represents the range
of model results for different disk structures. The three red
points represent the only known HD detections from Herschel-
PACS. \SALT\ SAFARI-Lite will make observations for
the full range of fluxes plotted, with 5$\sigma$, 1-hr sensitivities
between 0.5 and 2\ee{-20} W m$^{-2}$ across the 4 bands.
SALTUS HiRX will be sensitive enough to provide high
spectral resolution observations, with 5$\sigma$, 1-hr sensitivities
$> 5\ee{-19}$ W m$^{-2}$ in J=2-1 and $> 2\ee{-19}$ W m$^{-2}$ in J=1-0. Figure modified from \cite{Trap2017}.} 
\end{figure}

There is currently no observatory capable
of detecting HD. Near the end of its lifetime,
Herschel targeted HD in seven massive
disk systems, resulting in three detections
\cite{Bergin2013,McClure2016}, with HD-derived disk gas masses of
30-210 \mjup\ \cite{Calahan2021}. Crucially, these mass measurements
revealed that both CO and \water\ gas are
depleted in these disks relative to the ISM
\cite{Du2015,Schwarz2016}. \SALT\ SAFARI-Lite will measure
the total gas mass in hundreds of protoplanetary
systems over its 5-year baseline mission,
down to masses as low as 0.1 \mjup\ (Figure~\ref{fig:trapman}). When combined with observations of cold water
vapor, this determines the
amount of water removed from the outer disk
and transformed into water ice in the planet-forming
midplane \cite{Schwarz2023}.

Converting the HD detections into an accurate
total gas mass requires knowledge of the
disk temperature structure, as HD does not emit
appreciably below 20~K. The \SALT\ design
allows for full spectral coverage with SAFARI-Lite
or simultaneous observations in the four
HiRX bands. While integrating on the HD 1-0
and 2-1 lines in HiRX-3,4b, \SALT\ is able to
observe multiple optically thick \water\ and CO lines in
HiRX-1,2 spanning 55-1729 K in excitation energy,
compared to 128.49 K for the HD J=1 excited
state. These optically thick lines provide
direct measurements of the gas temperature
throughout the disk. 

\begin{figure}
\begin{center}
\begin{tabular}{c}
\hspace{-3mm}\includegraphics[height=8.0cm]{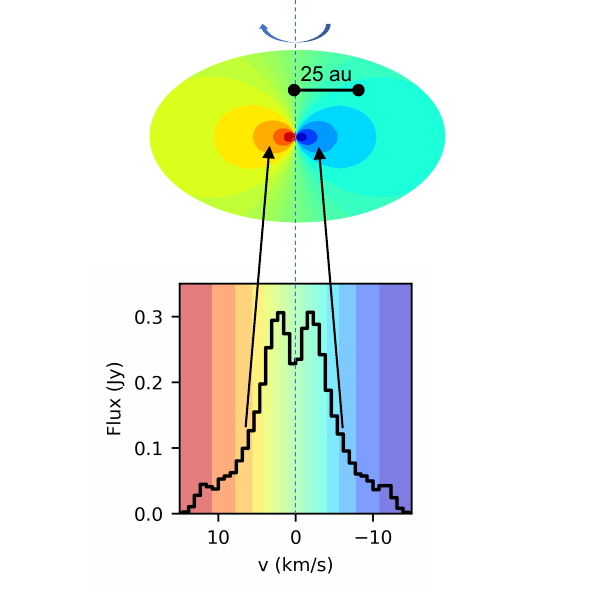} 
\end{tabular}
\end{center}
\caption 
{\label{fig:doppler}A simulated disk-integrated spectrum of HD 1-0 for a \eten{-3} \msun\ disk at a distance of 140 pc, inclined by 45 degrees. The spectral resolution is $\Delta v$ = 0.9 \kms. Disk emission is Doppler shifted due to
Keplerian rotation. Emission with a high velocity
offset from the line center originates from small radii,
while emission close to the systemic velocity
originates from large radii. The HiRX
high spectral resolution enables differentiation
between emission from different
radii in protoplanetary disks.} 
\end{figure}

The high spectral resolution of HiRX can then be used to map emission to different physical locations in the disk using a technique known as Doppler tomography or tomographic mapping.
Because disk rotation
follows a Keplerian velocity profile, the radius
at which gas emission originates can be determined
from the line profile. Thus, high spectral
resolution observations of molecular lines
in disks can be used to determine the radial location
of the emission without having to spatially
resolve the disk.
As shown in Figure~\ref{fig:doppler}, the velocity offset for
emission originating in the inner disk is of order
several \kms\ assuming a disk inclination of 45 degrees,
while in the outer disk the velocity offset is
much smaller. The velocity resolution ($\Delta v$) of
SALTUS HiRX is $< 1$ \kms, sufficient to distinguish
emission originating in the inner versus
outer disk.

Taking the expected HD fluxes and line-to-continuum
values into account from Figure~\ref{fig:trapman}
\cite{Trap2017}, SAFARI-Lite measures the J=1-0 and 2-1
lines at the 5$\mathrm{\sigma}$ level in 1 hour for the limits provided
in Figure~\ref{fig:trapman}, enabling reliable disk gas mass estimates. 
For a survey of disk mass across systems, which requires only the total HD flux,
spectrally unresolved observations with SAFARI-Lite
are able to quickly build a catalog of HD detections.
The expected continuum flux from a $3\ee{-5}$~\msun\ disk
at 140 pc, where many young stars are found, is 0.02 Jy \cite{Trap2017}. SNR of 300 requires a sensitivity of 66 $\mathrm{\mu}$Jy at 112\um, the wavelength of the HD 1-0 transition. Based on the modeled grating sensitivity of SAFARI-Lite (Roelfsema et al., this issue), this can be achieved in less than an hour on source. 
SAFARI-Lite’s greater sensitivity at 54 \um, the wavelength of the HD 2-1 transition, achieves SNR 300 in even less time.

For a subset of the brightest disks,
HiRX spectrally resolves the strong HD lines at
a ~1 \kms\ velocity resolution to measure the line
profile in detail and use Doppler tomography to
constrain the disk structures. As an example, TW Hya has a peak
flux of 0.49 Jy. \SALT\ HiRX Band 3 yields a
$5\sigma$ detection at $\Delta v = 1$ \kms\ in 20 hours. We can
expect to observe five targets per year in the
tomographic mode if we allocate 100 hours per
year to these observations. 
These deep HiRX observations of sources spanning several arcseconds on the sky will also provide constraints on the spatial extent of the line emission for these disks, important for validating the models used for interpretation of surveys.
In total, \SALT\ will
obtain the disk gas masses in hundreds of
protoplanetary systems during its nominal five year mission without the need for
ancillary data.

\subsection{The Spatial Distribution of Water in Protoplanetary Disks}\label{sec:diskwater} 

SALTUS will be the first mission with the sensitivity
to measure the distribution and physical
properties of water in a large sample of protoplanetary
disks. These measurements are key
for understanding planet formation and how terrestrial
planets acquire water. 
The \SALT\ instruments are designed to probe both the gas and the solid \water\ reservoirs and relate them to the characteristics of the central protostar (luminosity, spectral type) and of the
planet-forming disk (evolutionary state, mass, structure, temperature). The large frequency range
of HiRX and SAFARI-Lite provide access to many \water\ lines with a wide range of excitation
energies, tracing the cold-to-warm water vapor in disks, addressing Decadal
Questions E-Q1c, E-Q3b, and F-Q4b.

The first part of this program focuses on water vapor.
\emph{Herschel} revealed tantalizing but
tentative and limited evidence of water removal
from the surface layers of outer disks \cite{Hogerheijde11,Du2017}. \emph{SALTUS’s}
large improvement in sensitivity relative
to \emph{Herschel} makes observations of water in
disks routine and enables a complete survey of
water in all protoplanetary disks within 200 pc.
This large number of observations will allow \SALT\ users to
conclusively identify trends between the distribution
of water in disks and other properties,
e.g., dust disk size and the presence of substructure
\cite{Banzatti2020}.

\begin{figure}
\begin{center}
\begin{tabular}{c}
\hspace{-3mm}\includegraphics[height=10.0cm]{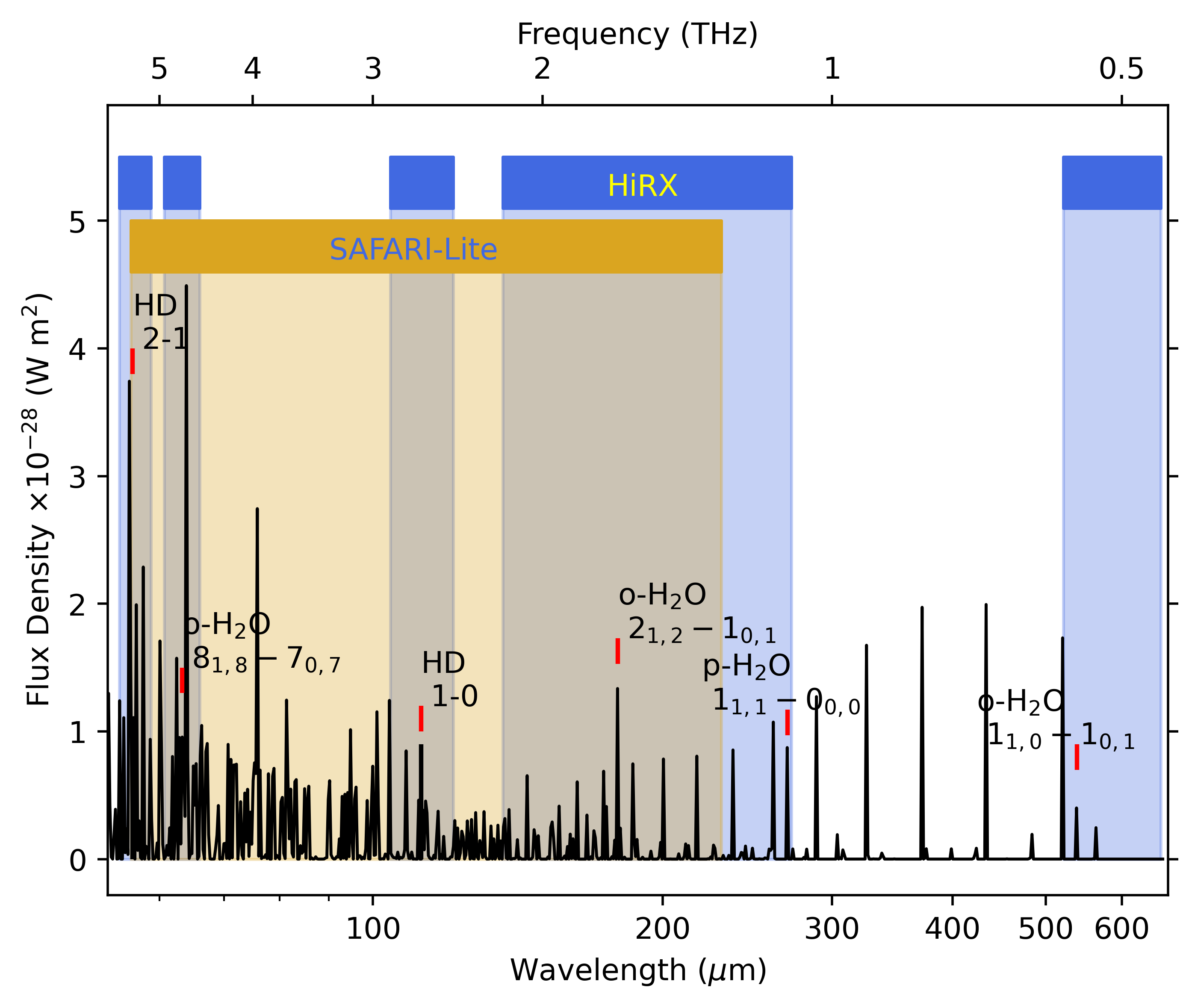} 
\end{tabular}
\end{center}
\caption 
{\label{fig:waterspectrum} The HiRX and SAFARI-Lite instruments will measure the gaseous emission spectrum of planet-forming disks, thereby determining unambiguously the disk gas mass, the water content, and the HDO/\water abundance ratios; all critical aspects of disk/planet formation models. The far-IR provides access to low energy transitions that probe the cold, outer parts of the disk of the planet-forming reservoir.} 
\end{figure}

Question E-Q1c from the 2020 Decedal asks: ``How common
is planetary migration, how does it affect
the rest of the planetary system, and what are
the observable signatures?'' 
The composition of a planet’s atmosphere is related to the composition of the disk
where and when it accreted its material and can be used to determine if a planet could have formed at its current
location or must have migrated. The chemical composition of the disk at a given location evolves over time
due to both chemical and dynamical processes \cite{Molliere22}. Current observational studies aim to use
atmospheric C/O to differentiate between early and late migration of Hot Jupiters \cite{Molliere22}. Models of
how migration changes C/O in a planet’s atmosphere make simplifying assumptions for C/O in
the disk \cite{Oberg2011}. The main volatile oxygen reservoir in disks, water, is virtually unconstrained by observations. \emph{JWST} is already significantly improving our understanding of water in
the mid-IR \cite{Perotti2023,Gasman2023,Sturm2023d,Banzatti2023b}. However, as noted by Decadal Question E-Q1c, additional longer wavelength
observations of cooler regions of the disk are needed to understand disk composition. HiRX will
map the radial distribution of cold water vapor in hundreds of protoplanetary disks. These disks will span a wide range of stellar mass and mass
accretion rate, disk dust mass, and disk radial extent, and span multiple starforming regions,
covering a variety of evolutionary stages \cite{Manara23}
HiRX will observe
the cold water vapor (not probed by
\emph{JWST}) by targeting the ground state ortho and
para transitions (Figure~\ref{fig:waterspectrum}), allowing users to
collect statistics on the cold water abundance in
disks across evolutionary stages. \SALT\ will provide the disk water measurements
needed to connect \emph{JWST} observations of exoplanet atmospheres to a formation time and location.

\SALT\ will tomographically map the water vapor distribution
toward a wide variety of disks, answering the
question “Where is the water?” Mapping the location
of water in protoplanetary disks is crucial
for understanding the transport of water during
planet formation \cite{roadmap}. Current planet formation
models predict that most planets form beyond
the snowline (gas/ice giants and possibly
smaller planets) and then experience
radial migration and dynamical scattering \cite{Lin1996}.
Confirming this prediction requires observational
constraints on the water snowline location
as functions of e.g., disk mass, stellar mass,
and evolutionary state, to compare to the orbital
radii of exoplanets in mature systems.

The ice snowline, the \water\ desorption
front located at $\sim$150-170 K in the disk,
controls the radial distribution of the C/O ratio
in the gas and solid phase \cite{Oberg2011}, implying that
the spectral characteristics of planets are linked
to their formation location. Water on terrestrial
planets which formed within the snow line (including
Earth) is thought to have been at least partially delivered by comet and asteroid impacts originating from
cold disk reservoirs beyond the snow line \cite{Raymond2017}. As
the mass of solids is expected to be the largest
near the snow line, the formation of giant planets,
such as Jupiter, is generally linked to the location
of the snow line in the solar nebula. Giant
planet formation may be aided by the increased
``stickiness'' of \water\ ice grains relative to minerals,
which greatly enhances the coagulation of
small dust grains \cite{Orme2009} – the first step in planet
formation – in the colder regions of these disks.

\SALT\ will probe the midplane water
snowline location by observing multiple high
upper-state energy ($E_u \sim 1000$ K) water lines,
which emit mostly from inside the midplane
water snowline \cite{Notsu2016}. 
Using tomographic mapping, \SALT\ will determine 
if water is returning to the gas with the
inward drift of icy dust grains, enriching the
water content of the terrestrial planet-forming
region. 
In contrast, \emph{JWST} primarily probes higher energy emission lines, with upper-state energies of several thousand Kelvin \cite{Banzatti2023b,Gasman2023}. Additionally, the dust continuum in the inner disk at mid-IR wavelengths is optically thin, such that the \water\ emission observable by \emph{JWST} originates in the photosphere. The continuum is less optically thick at the sub-millimeter wavelengths observed by \SALT, allowing us to probe deeper in the disk.
\SALT\ will make the first
measurements of the disk midplane water
snowline location in non-outbursting disks, an
important landmark in the core accretion picture.
\SALT\ enables us to assess the role of
the water snowline in determining the architecture of
planetary systems \cite{Kennedy2008,Fernandes2019} and the extent
to which processes (e.g., migration, dynamical
scattering) alter exoplanetary orbital radii; this
addresses Decadal Question E-Q1c.

\begin{figure}
\begin{center}
\includegraphics[height=8.0cm]{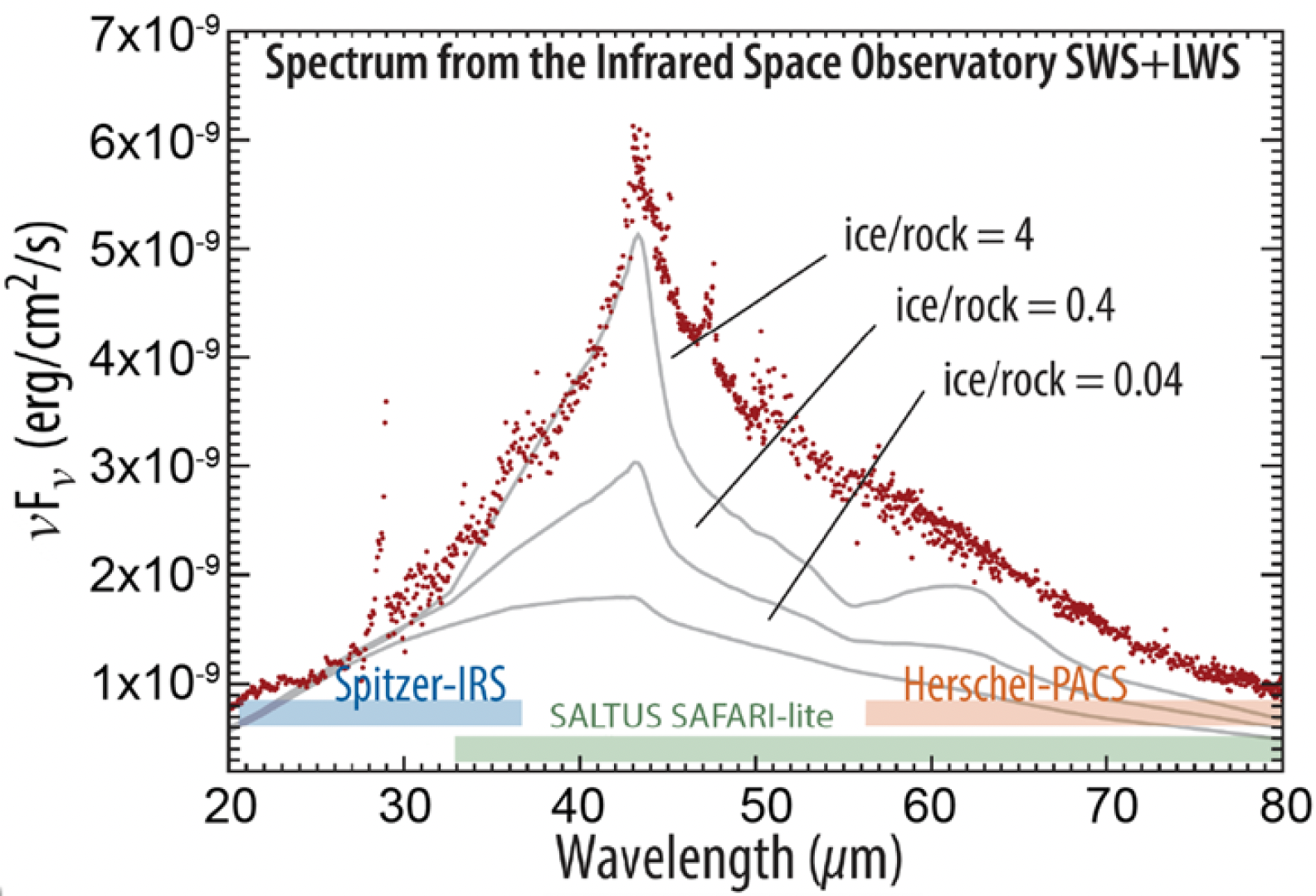}
\caption{The far-IR
spectrum of the Herbig star, HD
142527, measured by
ISO/SWS+LWS and Herschel/
PACS (red points), reveals the presence
of the lattice modes of crystalline
\water\ ice at ~43 and 62 \micron\ \cite{Min16}. This is compared to emission models at different ice/rock ratios (grey lines). The tell-tale signature of crystalline water ice is recognizable to ice/rock ratios as small as 0.1. For
comparison, the estimated ice/rock ratio in the solar nebula during planet formation is 1-2.}
    \label{fig:ice}
\end{center}
\end{figure}

The second part of the program will target water ice directly. 
Water ice is the most abundant non-refractory
solid-state component of planet-forming
disks, locking up a major fraction of the elemental
oxygen.
The water ice distribution in protoplanetary
disks is of fundamental importance for our
understanding of planet formation and their
characteristics.
As a result of its simultaneous spectral coverage of the full 34-230 \micron\ range and its high sensitivity, SAFARI-Lite is uniquely suited to study emission in the the diagnostic lattice modes of ices in protoplanetary disks (Figure~\ref{fig:ice}), providing temperature, mass, and structure of the emitting ices. Previous far-IR space missions (S\emph{pitzer, Herschel, ISO}) lacked the wavelength coverage or sensitivity for a systematic study of far-IR ices, especially in planet-forming disks. While the NIRSpec and MIRI instruments on \emph{JWST} cover the near- and mid-IR region, home to ice fundamental modes, these shorter wavelengths require a very favorable viewing angle – almost edge-on – and cannot perform a systematic study of the role of ices in planet-forming disks. 
Further, because these features are seen in absorption, they provide only a lower limit on the absorbing column, as a photon's path as it is scattered through the disk is uncertain \cite{Sturm2023c,Sturm2023d}. The far-IR features have the advantage of being seen in emission, and are therefore not subject to same constraints due to viewing angle and scattering.

The large wavelength coverage and moderate spectral resolution of SAFARI-Lite are
well matched to the expected profile variations in the lattice modes of \water\ ice, measuring the
temperature history of the ice grains. This is linked to a physical location through models of disk
temperature structure, constrained by the \water\ gas observations. The gas and solid reservoirs
interact through sublimation and condensation as icy grains drift inwards from the cold outer disk
to the warm inner disk and through turbulent cycling between the colder mid-plane and the warmer
disk photosphere. \SALT\ will quantify the mass of the gaseous \water\ and
ice reservoirs in a large sample of protostellar and protoplanetary sources, assess the interrelationship of these
reservoirs, and connect them to physical characteristics of the stars and their disks and thereby
address the importance of the physical processes that link them.

\subsection{Water in Prestellar Cores}\label{sec:molclouds} 
Prestellar cores are the gravitationally bound phase of star formation immediately prior to the protostar formation\cite{Bergin2007,Andre2014,Pineda2023}, with cold (T$<10$ K), dense ($n>10^5$ cm$^\mathrm{-3}$) central regions that are well shielded from the surrounding interstellar radiation field. During this phase, the initial chemical conditions are set for the disk and subsequent planet formation. The direct chemical inheritance from the prestellar phase to the protostellar disk has been established, e.g., reflected in the D/H ratio from ALMA observations of deuterated water \cite{Jensen2021}. 

Although most of the water in prestellar cores resides in the solid state on the dust grain icy surfaces \cite{Bergin2002}, photodesorption by UV photons can liberate water molecules into the gas phase at abundances that are typically $<\eten{-9}$ with respect to \hh\ \cite{vanDishoeck2021}. Two main sources of UV photons exist: the surrounding interstellar radiation field is the dominant heating component of dust grains \cite{Evans2001} and a low intensity UV radiation field from \hh\ excitation due to collisions with electrons that come from cosmic ray ionizations of \hh\ and He \cite{Prasad1983}.

The $1_{10} - 1_{01}$ ground state rotational transition of ortho-\water\ at 538.2 \micron\ (557 GHz; HiRX 1) can be observed in absorption against the continuum of the prestellar core \cite{vanDishoeck2021}. The line can also be seen in emission if the central density of the prestellar cores is $>10^7$ cm$^\mathrm{-3}$, although only a few prestellar cores are known that have this extreme central density \cite{Caselli2012}. The gas phase water in the outer part of the core at low A$_\mathrm{V}$ has a photodesorption rate that depends on the strength of the interstellar radiation field ($G_0$), and a constraint on $G_0$ is needed to determine the dust temperature and the gas temperature profiles in the outer part of prestellar cores \cite{Young2004}. Accurate temperature profiles are crucial for radiative transfer modeling of molecular emission and absorption observed toward prestellar cores, and \SALT\ water vapor observations of prestellar cores will play an important role in constraining the temperature profile in the outer part of the cores.

\subsection{Astrochemistry: CHONS from cores to disks} \label{sec:astrochem} 
Hot cores are hot molecular line emission regions within massive star-forming regions, typically characterized by high temperatures (100s of K) and densities ($\sim$\eten{7} cm$^\mathrm{-2}$) \cite{vanderTak2000}. Originally identified from the detection of hot \ammonia\ towards Orion-KL \cite{Morris1980}, hot cores were subsequently found to host an incredibly rich gas-phase organic chemistry \cite{Blake1987}. Ice mantles are the main sites of astrochemical complex organic molecule formation, and ice sublimation is the source of the chemical complexity detected in hot cores \cite{Garrod2006,Herbst2009}. Observing molecular line emission from hot cores provides powerful constraints on their physical and chemical conditions \cite{Garrod2013}. 

\begin{figure}
\begin{center}
\begin{tabular}{c}
\hspace{-3mm}\includegraphics[height=5.0cm]{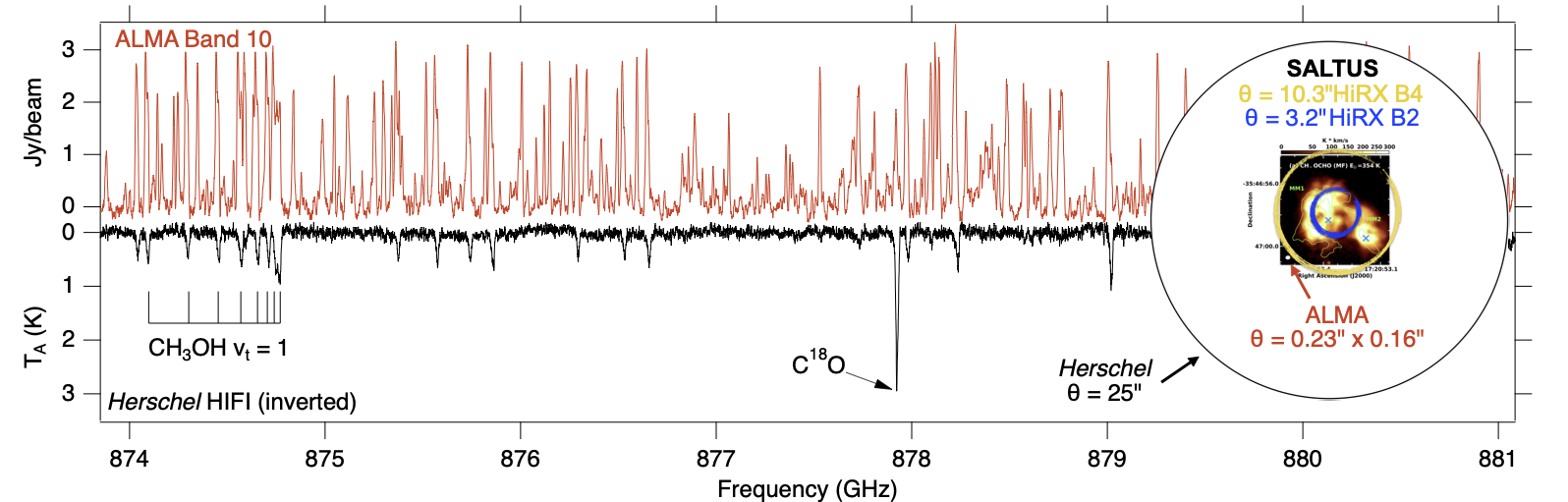} 
\end{tabular}
\end{center}
\caption 
{\label{fig:comscompare} Comparison of observations of the high-mass star formation region NGC 6334I taken with ALMA (red) and Herschel (black and inverted). The inset shows the beam size for Herschel, ALMA, and \SALT\ overlayed on a methyl formate intensity map of the region. The effects of beam dilution will be much less severe with SALTUS than with Herschel. Figure adapted from \cite{McGuire2018}.} 
\end{figure}

\emph{SALTUS's} high sensitivity at far-IR wavelengths will open a new window into studying complex organic molecules in hot cores. Figure~\ref{fig:comscompare} illustrates how a massive star-forming region can appear line-poor with \emph{Herschel} but harbor hundreds of spectral lines when observed with a higher sensitivity and resolution observatory (ALMA Band 10). With \SALT, we similarly expect higher line densities of organics compared to Herschel. While the sensitivity increase with \SALT\ will be more modest than with ALMA, we note that ALMA Band 9 and 10 observations require exceptional weather conditions and do not extend to wavelengths shortward of 315 \um, whereas \SALT\ will provide access to wavelengths as short as 34 \um.
 
While many complex organics can be detected at longer wavelengths, there are several advantages to obtaining far-IR observations. First, the lines covered by \SALT\ typically probe higher upper state energies than millimeter-wavelength lines, which can better constrain excitation conditions. This is especially important for high-mass hot cores, in which organics often have excitation temperatures of a few 100 K \cite{Crockett2014b,Neill2014}. Constraints on organic molecule excitation temperatures are required to interpret the physical conditions of the emitting regions, as well as the chemical relationships between different classes of molecules, see also\cite{Bergner2022}.

The early Class 0 and I stages of low-mass protostellar evolution, characterized by an infalling envelope of gas and dust, are often accompanied by an outflow, which promotes accretion onto the protostar by carrying away angular momentum. Encounters between the outflow and the ambient envelope material produce shocks, which can alter the local chemistry through heating and  grain sputtering. In some “chemically rich” outflows, the gas-phase abundances of molecules associated with the ice phase (H$_{\mathrm{2}}$CO, CH$_{\mathrm{3}}$OH, CH$_{\mathrm{3}}$OCHO) are enhanced due to shock-induced ice sputtering \cite{Garay1998,Codella1999,Requena2007,Arce2008}. Thus, these chemically rich outflows offer a valuable window to probe the organic composition of interstellar ices. Moreover, studies of outflow shock physics and chemistry inform our understanding of the same processes that take place on smaller, disk-forming scales within the protostellar core.

\begin{figure}
\begin{center}
\begin{tabular}{c}
\hspace{-3mm}\includegraphics[height=8.0cm]{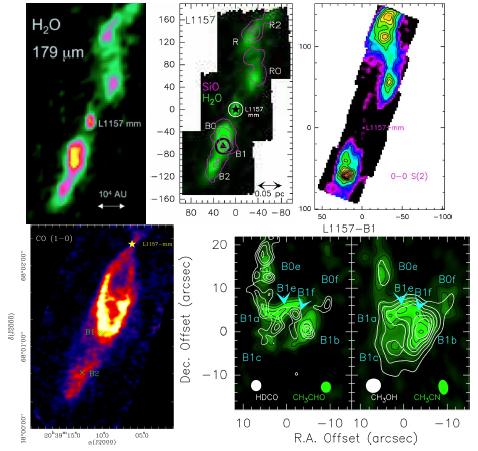} 
\end{tabular}
\end{center}
\caption 
{\label{fig:shocks} Observations of an outflow from a low mass protostar. 
Top left: \water\ distribution as measured by \emph{Herschel} in the $2_{12}-1_{01}$ transition.
Top center: SiO contours overlayed on the \water map. Top right: Pure rotational \hh\ emission from the $S(2)$ transition. Bottom left: Blue-shifted outflow lobe in CO J=2-1. 
Bottom right: Blow up of the B1 shock at $1\ensuremath{''}$ spatial resolution in acetaldehyde with HDCO contours (left) and methyl cyanide with methanol contours (right). 
At high spatial resolution the outflow resolved into many small dense clumps with very distinct and highly variable chemical composition. 
Figure from \cite{Tielens21}. } 
\end{figure}

 The archetypical chemically rich outflow shock, L1157-B1, was observed as part of the \emph{Herschel} CHESS survey \cite{Ceccarelli2010}. The 471-540 \micron\ spectrum contained emission lines from high-excitation transitions of grain chemistry tracers like \ammonia, H$_{\mathrm{2}}$CO, and CH$_{\mathrm{3}}$OH \cite{Codella2010}. An excitation analysis revealed that these lines emit with temperatures $\geq 200$ K, intermediate between the cold emission observed by longer-wavelength transitions and the very hot gas traced by \hh\ emission. Thus, observations of higher-excitation organics towards outflow shocks can help link these different emission regimes and disentangle how the shock chemistry and physics progresses (Figure~\ref{fig:shocks}). These insights can in turn be used to refine models of shock astrochemistry, which are needed to connect observed gas-phase abundances to the underlying grain compositions \cite{Burkhardt2019}.

Lastly, chemically rich outflow shocks are the only low-mass star forming regions where phosphorus carriers have been detected \cite{Yamaguchi2011,Bergner2019c}. In shock chemistry models, PH$_\mathrm{3}$ and smaller P-bearing hydrides are predicted to be at least as abundant as the P carriers PN and PO \cite{Jimenez2018}. PH$_\mathrm{3}$ has only one strong transition observable longward of 600 \micron, and remains undetected in star-forming regions. \emph{SALTUS's} broad spectral coverage measurements allow for a more complete inventory of the volatile phosphorus carriers in star-forming regions.

\subsubsection{Astrochemistry: CHONS in disks} \label{sec:diskastrochem} 
While molecules observable at millimeter wavelengths have been extensively studied in disks, there are almost no constraints on the inventories of light hydrides in disks, many of which are observable only at submillimeter/far-IR wavelengths. Perhaps the most exciting observations of light hydrides enabled by \SALT\ are observations of \ammonia. Indeed, the N budget in disks is poorly constrained given that the dominant N carrier, \nn, cannot be directly observed in the gas. Ice spectroscopy towards low-mass protostars, the evolutionary progenitors of disks, has revealed that \ammonia\ is an important N carrier in the ice, with relative abundances of $\sim 5\%$ with respect to \water\ compared to $<1\%$ in nitriles, or XCN \cite{Oberg2011a}. While nitriles are commonly detected towards disks \cite{Dutrey1997,Oberg2015,Guzman2017,Bergner2019b,vanTerwisga2019b}, \ammonia\ has only been detected towards two disks. The 524.1 \micron\ transition of o-NH$_\mathrm{3}$ was first detected by Herschel towards the nearby TW Hya disk \cite{Salinas2016}, and \ammonia\ was also detected towards the embedded (Class I) disk GV Tau N at mid-IR wavelengths tracing hot emission from the inner few au \cite{Najita2021}. HiRX Bands 1 and 2 will cover multiple strong transitions tracing cool \ammonia\ (upper state energies 27-170 K).

\SALT\ observations of multiple \ammonia\ lines will allow for the first \ammonia\ excitation analysis in the outer disk. Additionally, \emph{SALTUS's} high spectral resolution will enable a kinematic analysis of the \ammonia\ line profiles in sources with high SNR, providing constraints on the spatial origin of the emission and the location of the \ammonia\ snowline. Auxiliary constraints on the disk structures, provided by \SALT\ observations of CO isotopologues, HD, and \water, will permit robust \ammonia\ abundance retrievals. The \ammonia/\water\ abundance ratio is of particular interest, as it can be directly compared with the ratio measured in comets to provide insights into how N is inherited by solar system bodies.

Another promising avenue for disk science with \SALT\ is S-bearing hydrides. Sulfur is commonly very depleted from the gas in dense star-forming regions, though several S carriers (CS, SO, H$_\mathrm{2}$S, H$_\mathrm{2}$CS) have been detected in disks \cite{Dutrey2011,Guilloteau2013,LeGal2019}. H$_\mathrm{2}$S was only recently detected in Class II disks: first towards GG Tau A \cite{Phuong2018}, followed by UY Aur and AB Aur\cite{Riviere2021,Riviere2022}. Towards other well-known disks, deep searches for H$_\mathrm{2}$S have only produced upper limits \cite{Dutrey2011}. To date, only the $1_{10}-1_{01}$ line at 168.73 GHz has been targeted, which is readily observable by ground-based telescopes but also intrinsically weak compared to the higher-frequency lines covered by SALTUS. The H$_\mathrm{2}$S lines at 160.7 \micron\ and 233.9 \micron\ appear particularly promising for detection in disks with \SALT, particularly if the emission originates in a somewhat warm environment.

In addition to the \water\ ice phonon modes discussed above, SAFARI-Lite's broadband coverage 
spanning 30 to 230 \micron\ will cover unique spectral signatures from a large number of volatile 
ice species, most notably \nn, O$_\mathrm{2}$, \cotwo, CO, \methanol, \methane, H$_\mathrm{2}$S,
\ammonia, and HCN. The uniqueness of the lattice modes enables us to clearly distinguish between 
the amorphous and crystalline ice phases, opening up a window to phase transition temperatures, 
which ultimately informs on the thermal evolution of the ice. Ice lattice modes are also the best 
viable way to determine the presence of homo-nuclear molecules such as O$_\mathrm{2}$ and \nn,
whose fundamental modes are IR inactive. The possibility to quantify the abundance of \nn\ ice in 
protoplanetary disks is particularly interesting as \nn\ is likely a major carrier of nitrogen \cite{Schwarz2014,Pontoppidan2019}.

\subsection{D/H Ratios as a Probe of Interstellar Heritage} \label{sect:DH} 
\begin{figure}
\begin{center}
\begin{tabular}{c}
\hspace{-3mm}\includegraphics[height=6.75cm]{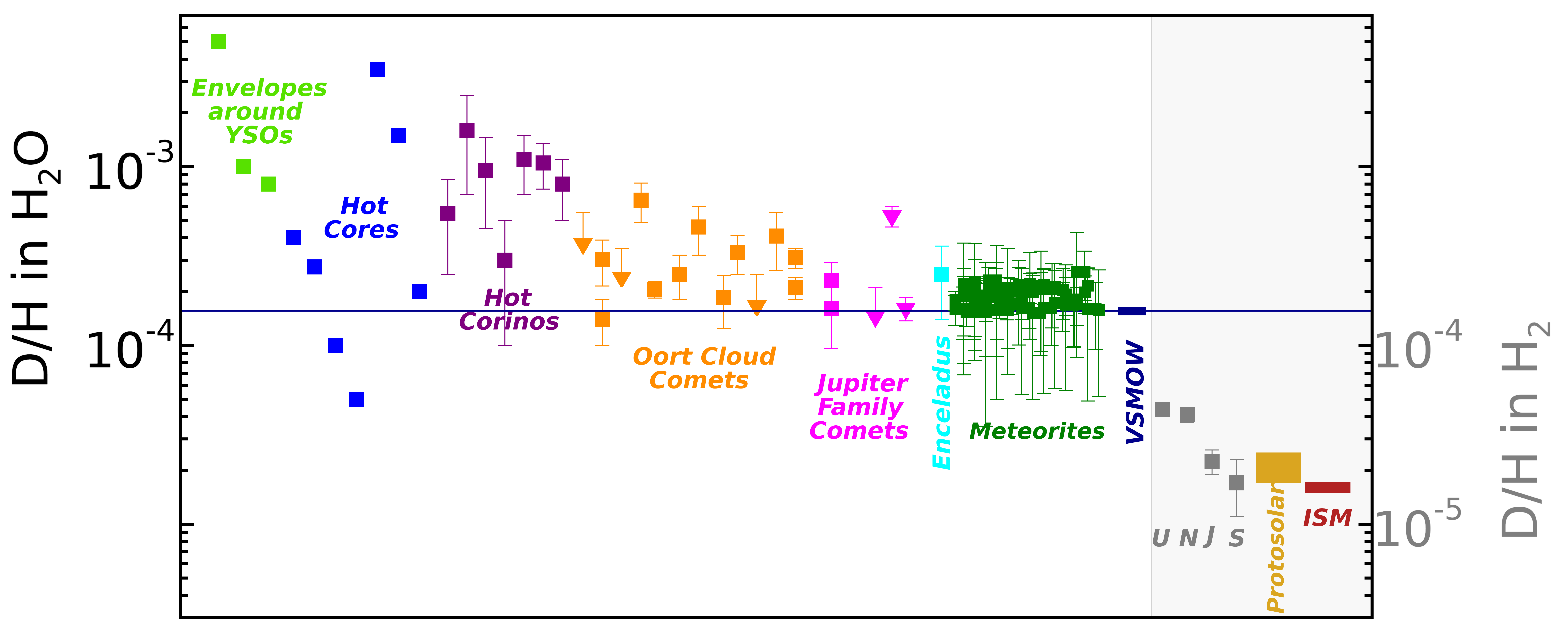}
\end{tabular}
\end{center}
\caption 
{\label{fig:DH} Measured D/H ratios in various Galactic and solar system sources. D/H values are based on measurements of HDO and \water\ (left axis) for all objects with the exception of the solar system gas giants, the protosolar nebula, and the ISM, where the D/H values are based on HD/\hh. D/H from HD/\hh\ trace the dominant reservoir of deuterium, while \water\ based measurements provide insight into the chemical history of \water. The D/H ratio of Earth's water, as measured by VSMOW is shown by the navy rectangle.  
Figure patterned after \cite{Anderson2022}, in turn adapted from\cite{Hart2011b,Lis2013,Bock2012}. 
Values are taken from \cite{Jens2019,Bive2016,Vill2009,Paga2017,Lis2019,Cout2012,Cout2013,Cout2014,Pers2014,Wang2012,Empr2013,vander2006,Helm1996,vanDishoeck2021,Bona2013,Yang2013,Jacq2013,Gibb2016,Bive2024}.
}
\end{figure}

Water is a key ingredient in the emergence of life and is, therefore, a key aspect in the assessment of the habitability of (exo)planets. Yet, the origin and delivery of water to habitable planets and notably Earth remains unclear. Terrestrial water could have been delivered by water-rich asteroids driven by the migration of Jupiter in the solar nebula and/or by the late heavy bombardment during a solar system-wide rearrangement \cite{OBrien2014,Raymond2017}. Outgassing from the deep mantel likely also contributed to Earth's surface water \cite{Broadley2022}. The enhanced D/H ratio in standard mean ocean water (SMOW) of 1.5\ee{-4} \cite{Hagemann1970} relative to the interstellar elemental D/H ratio [1.5\ee{-5}; \cite{Prodanovic2010}] provides support for this view as deuterium fractionation is a chemical signature indicating that a fraction of water formed under cold conditions, likely at the surface of interstellar grains (Figure~\ref{fig:DH}) \cite{Tielens1983}. 

This anomaly would reflect the effects of chemistry at low temperatures in cold prestellar cores where the small zero-point energy difference between D- and H-bearing species can create large deuterium fractionations \cite{Tielens1983,Ceccarelli2014}. However, the observed D/H ratio in deeply embedded protostars (hot corinos) – tracing the inherited water content – is higher than the D/H ratio in Earth's water (VSMOW) by factors of ~2 to 6 (purple symbols in Figure~\ref{fig:DH}). Hence, chemical processing must have occurred in warm gas, reducing the deuterium fractionation. Likely, this reprocessing of the water occurred in the warm surface layers of protoplanetary disks – on a disk-wide scale – where radiation from the young star photo-desorbs \water\ from preexisting ices and reforms water through gas phase reactions. The variation in measured D/H ratios for various astronomical objects provides important clues to the formation conditions at different locations in nascent planetary systems. \SALT\ will help to unravel the following questions: What is the HDO/\water\ ratio in protoplanetary disks and how does that depend on the characteristics of the protostar, the conditions in the protoplanetary disk, and the molecular core environment? What processes play a role in the water cycle of protoplanetary disks? 

\SALT\ will detect deuterated isotopologues of complex organics in hot corino and non-hot corino sources. While a 1-hour HiRX integration will provide $>5\sigma$ detections of single deuterated $\mathrm{CH_3OCH_3}$, double deuterated $\mathrm{CH_3OCH_3}$, and single deuterated $\mathrm{C_2H_5OH}$ in hot corinos, a 10-minute integration will yield a robust detection of the strongest lines of deuterated molecules in protostellar envelopes.

\begin{figure}
\begin{center}
\begin{tabular}{c}
\hspace{-3mm}\includegraphics[height=10.0cm]{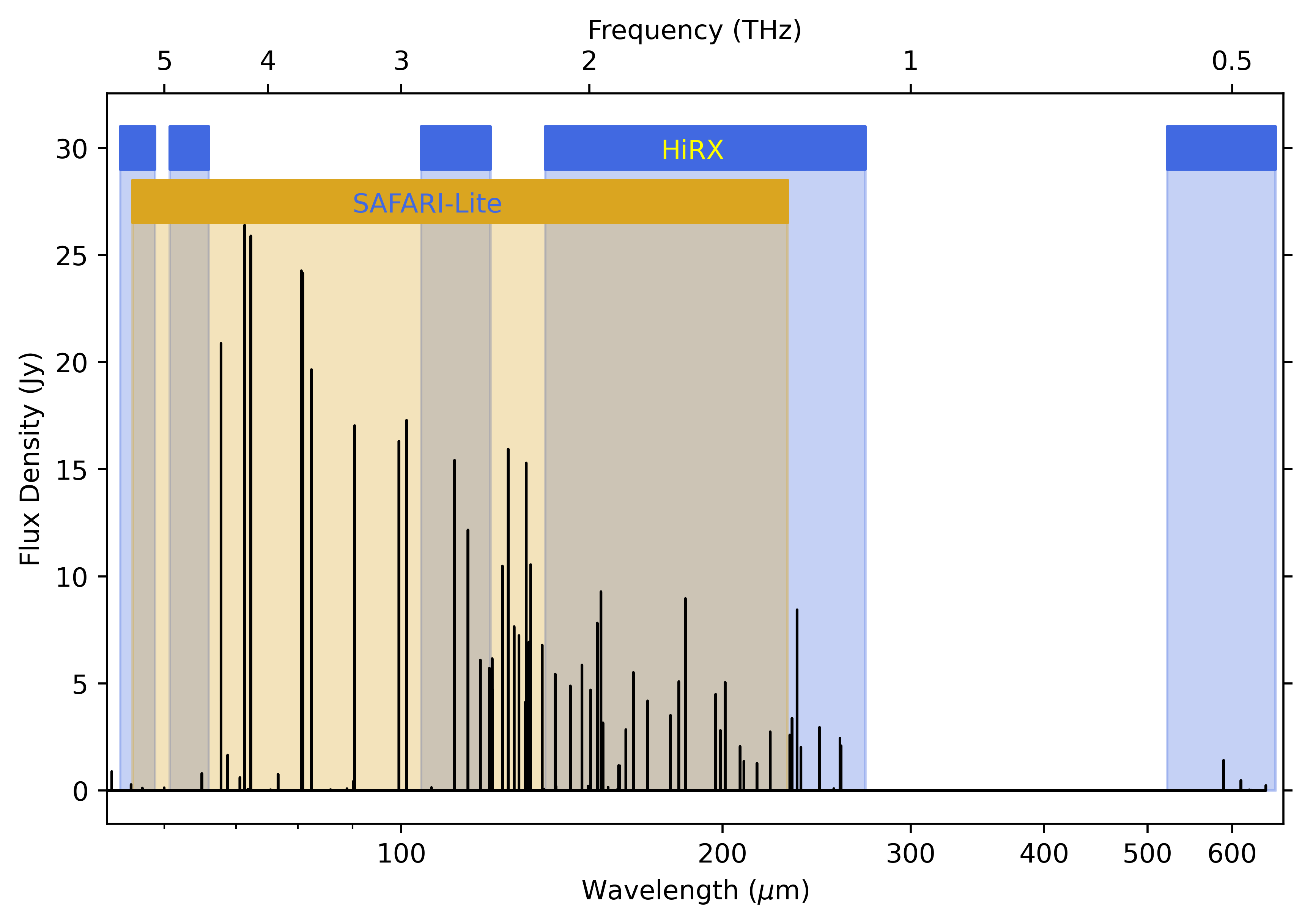}
\end{tabular}
\end{center}
\caption 
{\label{fig:diskHDO} Predicted HDO line emission in TW Hya, based on the best fit physcial/chemical disk model of \cite{Calahan2021} and assuming HDO/\water = 2\ee{-3}.}
\end{figure}

The D/H ratio in protoplanetary disks has been probed primarily through trace species such as DCO$^+$ and DCN \cite{Huang2017,Salinas2017,Munoz2023}. 
ALMA observations have constrained the D/H ratio in water for one disk, V883 Ori, based on detections of HDO and H$\mathrm{_2^{18}}$O at 200 GHz \cite{Tobin2023}. 
In this system, the D/H ratio was found to be similar to that for water in the ISM. However, V883 Ori is an exceptionally warm disk currently undergoing an accretion burst, thus increasing the observable water column. The main isotopologues of water are difficult to observe from ground-based facilities, even at high altitude \cite{Facchini2024}. 
Such observatories are often limited to the much weaker H$\mathrm{_2^{18}}$O lines, which are still impacted by the low atmospheric transmission at the relevant frequencies \cite{vander2006}.
We identify the strongest transitions of HDO using the physical/chemical disk model of \cite{Calahan2021} (Figure~\ref{fig:diskHDO}). 
This model of the nearby disk TW Hya reproduces the resolved ALMA observations of multiple CO transitions as well as the total HD 1-0 flux from Herschel and the upper limits on the HDO 225 GHz line from the Submillimeter Array (SMA) \cite{Qi2008,Bergin2013}. 
The strongest HDO transitions in protoplanetary disks are at 71.4 \micron, and inaccessible from the ground. 
These observations will provide the link between water in the ISM to water in planetary systems, providing a definitive answer to whether water on terrestrial planets is commonly inherited from the ISM.

\subsection{Debris Disks} \label{sec:debrisdisks} 
The debris disk phase follows the protoplanetary
disk phase. Debris disks are gas-poor, with broad disks or
rings of second-generation dust thought to be
influenced by the presence of planets \cite{Pearce22}. Debris
disk observations allow us to study populations
of small bodies around other stars and infer the
presence of planets that otherwise evade detection \cite{Lagrange10,Skaf23}.
They also provide insight into the
composition of solid bodies in other planetary
systems \cite{Mittal15,Rodigas15}.
Debris disks may play a role in planet formation because of the gas (with total masses up to $\sim$1 M$_\oplus$) that is now observed in these disks \cite{Moor2017}. Indeed, this gas could spread and accrete onto planets, thus changing their initial atmospheric compositions between 10-100 Myr \cite{Kral20}. This secondary gas component may also be important to understand our own Solar System and find out whether the Kuiper belt can still release gas today or whether it may have contained gas in its youth \cite{Kral2021}. If this gas could got accrete onto the giants, it may explain, e.g., the high metallicity of Uranus and Neptune \cite{Kral20}. 
\SALT\ has the potential to observe Kuiper Belt analogues, which is be necessary to be able to explore these questions.

\subsubsection{Kuiper Belt Analogues} \label{sec:exokuiperbelt} 
Debris disks with the same intrinsic luminosity
as the Solar System’s Kuiper Belt
have yet to be observed \cite{Eiroa13}. These exo-Kuiper
Belts have typical temperatures of $\sim$50 K, corresponding
to a black-body emission peak in the
far-IR. Updating sensitivity estimates from the
original \emph{SPICA} SAFARI \cite{Kral2017} to \emph{SALTUS’s}
SAFARI-Lite, \SALT\ will reach the $5\sigma$ sensitivity
threshold to detect exo-Kuiper belts
around the nearest 30 G and K stars with
known debris disks in 1 hour of integration.
\SALT\ can determine the frequency of exo-Kuiper Belts, characterizing how common dust
is in mature planetary systems, thus addressing
Decadal Question E-Q1d. Additionally, the angular
resolution of \SALT\ should enable
mapping of the dust in some of these systems, thereby serving as a Rosetta Stone between the
far-IR and longer wavelength observations of
ALMA. 
It is possible that the massive exo-Kuiper belts detected to date prevent the development of life in the habitable zone due to an excessively high bombardment rate. 
In this case, targeting systems with belts similar to ours (i.e. with low masses) could 
help optimize the search for life on another planet.

\subsubsection{Gas in Debris Disks} \label{sec:debrisdiskCO} 
\SALT\ can observe the gas content in debris disks, which are expected to
contain detectable levels of carbon and oxygen;
based on the gas production model developed
by \cite{Kral2017,Kral2019}. These observations, focusing on
ionized carbon and neutral oxygen, will complement
those made by ALMA, which targets
CO and neutral carbon \cite{Kral2016b,Cataldi2020}. By doing so,
\SALT\ can gather valuable information
about the carbon ionization fraction, a crucial
factor in understanding the dynamics of gas,
including determining the dominant mechanism of angular momentum transport.
Possibilities include 
the magneto-rotational instability (MRI) \cite{Kral2016a}, or MHD winds or even some hydrodynamic instabilities such as vertical shear instability (VSI) or Rossby Wave Instability (RWI) \cite{Cui24}. These different mechanisms operate at different ionization fractions, densities, and depend on the magnetic field configuration as well. Only new data for a large variety of systems will allow us to pinpoint the dominant mechanism. One can also use the spatial information (extracted from high spectral resolution) to rule out some mechanisms, as for instance, MHD winds are expected to only produce viscous expansion inwards. In contrast, turbulence will also allow the gas to extend further outwards than its production source.

The low surface density in debris disks allows penetration of
a high photon flux from the central star,
converting molecules like CO, \cotwo, and \water\
into ionized atomic carbon and oxygen via photodissociation and photoionization.
By targeting [CII] and [OI] in the far-IR, \SALT\ users gain insights into the initial
species released from planetesimals by examining
the C/O ratio, e.g., to investigate whether CO,
\cotwo, or \water\ is released, which could have strong connections with TNOs in the Solar System for which we can now probe the composition with the \emph{JWST} \cite{depra24}. These observations
provide a comprehensive understanding of the
gas disk composition at different radii from the
central star. The accretion of carbon and oxygen
by young planets may play a pivotal role in the
formation of the building blocks of life \cite{Sutherland2017,Pabs2021}
 or affect the temperature through
greenhouse effects, thus influencing their habitability.
Currently, debris disk studies have been
mainly with A stars \cite{Kamp2021}. \SALT\ has the
sensitivity to detect [CII] and [OI] in the more common
FGK stars. These observations determine
the C/O ratio across spectral type during the late
stages of planet formation, when volatile gasses
are delivered to terrestrial planets, and address
Decadal question E-Q3a “How are potentially
habitable environments formed?”
\SALT\ can particularly look at this question around solar mass stars.

\subsection{References}






\subsection*{Disclosures}
The authors have no relevant financial interests in the manuscript and no other potential conflicts of interest to disclose.

\subsection* {Code, Data, and Materials Availability} 
This paper reviewer the science cases and potential observations for a future space mission and so data sharing in not applicable at this time. 




\vspace{2ex}\noindent\textbf{Kamber R. Schwarz} holds a postdoctoral position at the Max Planck Institute for Astronomy in
Heidelberg. She was a NASA Sagan Postdoctoral Fellow at the Lunar and Planetary Laboratory
at the University of Arizona. She received a PhD in Astronomy \& Astrophysics at the University
of Michigan in 2018. She studies the evolution of volatile gas during planet formation, with the
goal of determining the amount of volatile carbon, nitrogen, and oxygen available to form
planets. Her research combines observations from the infrared to the millimeter, using facilities
such as ALMA, NOEMA, and JWST, with physical/chemical modeling to constrain the timescales
and mechanisms of volatile reprocessing. She has authored about 100 publications.

\vspace{2ex}\noindent\textbf{Alexander Tielens} is a professor of astronomy in the Astronomy Department of the University of Maryland, College Park. He received his MS and PhD in astronomy from Leiden University in 1982. He has authored over 500 papers in refereed journals and has written two textbooks on the interstellar medium. His scientific interests center on the physics and chemistry of the interstellar medium, in particular in regions of star and planet formation.

\vspace{2ex}\noindent\textbf{Joan Najita} is an Astronomer at NSF’s NOIRLab and its Head of Scientific Staff for User
Support. She was formerly the Chief Scientist at the National Optical Astronomy Observatory
(NOAO) and served on its scientific staff since 1998. In 1993 she received her PhD
from University of California, Berkeley. Najita has been responsible for strategic planning,
science career development, science communications, and the health of the scientific
environment at the Observatory. Her interests include traditional research topics (such as star and
planet formation, exoplanets, and the Milky Way), advocacy for the development of new
research capabilities (such as infrared spectroscopy and massively multiplexed wide-field
spectroscopy), as well as the sociological context of astronomy (such as the nature of discovery
in astronomy, and its science sociology and resource allocation practices). She has a lifelong
interest in communicating science to the public and in the role of science in society. Joan Najita
has been named a 2021–2022 fellow at Harvard Radcliffe Institute, joining artists, scientists,
scholars, and practitioners in a year of discovery and interdisciplinary exchange in
Cambridge. She has authored about 190 publications

\vspace{2ex}\noindent\textbf{Jennifer Bergner} is an Assistant Professor of Chemistry at UC Berkeley. She received her BS
degree from University of Virginia and MA and PhD from Harvard in 2019. Her astrochemistry
group uses a variety of tools to explore the chemistry at play in protostars and protoplanetary
disks, the progenitors of planetary systems. With cryogenic vacuum experiments she mimics the
extremely low temperatures and pressures of star-forming regions in the lab to explore the
chemical and microphysical behavior of volatile ices. She also uses state-of-the-art telescope
facilities like ALMA and JWST to observe the spectral fingerprints of volatile molecules in
protostars and protoplanetary disks, providing insight into the chemical landscape of planet
formation and the underlying physical processes which drive astrochemical evolution. She has
about 128 publications.

\vspace{2ex}\noindent\textbf{Quentin Kral} is an astronomer at the Paris Observatory (LESIA). His main research interests are debris disks, the solar system, and planetary formation. He is an expert on the new gas component that is now observed in mature extrasolar systems once the young planet-forming disk has dissipated. He mainly uses ALMA to test his models and investigate the gas and dust in exoplanetary systems. He is the PI of the exoplanet.eu catalog of exoplanets. He has published over 130 articles. He received his master's degree from Ecole Normale Supérieure (ENS Paris) and his PhD from Paris Observatory in 2014.

\vspace{2ex}\noindent\textbf{Carrie M. Anderson} is a research scientist at NASA Goddard Space Flight Center (GSFC). She received a BS in physics from Arizona State University in 2000, and MS and PhD degrees in Astronomy from New Mexico State University in 2003 and 2006, respectively. She is the author of more than 45 papers in refereed journals and has written one book chapter. Her research focuses on the remote sensing of planetary atmospheres, primarily in the areas of thermal structure and composition, using space- and ground-based data, in the visible, near-IR, mid-IR, far-IR, and submillimeter spectral regions. Her research also includes laboratory transmission spectroscopy measurements of ice films in a high-vacuum cryo chamber located in her Spectroscopy for Planetary ICes Environments (SPICE) laboratory at NASA GSFC.

\vspace{2ex}\noindent\textbf{Gordon Chin} is a research scientist at NASA Goddard Space Flight Center (GSFC). He received his B.A. in physics from Columbia College in 1970, and his M.A., M. Phil., and PhD in physics from Columbia University in 1972, 1974, and 1977, respectively. He is the author of more than 50 refereed journal papers. His current research interests includes the development of sub-millimeter planetary flight spectrometers targeting planetary atmospheres, the lunar exosphere, and ocean world plume environments in the solar system.

\vspace{2ex}\noindent\textbf{David T. Leisawitz} is an astrophysicist and Chief of the Science Proposal Support Office at
NASA’s Goddard Space Flight Center. He received a Ph.D. in Astronomy from the University of
Texas at Austin in 1985. His primary research interests are star and planetary system formation,
infrared astrophysics, wide-field spatio-spectral interferometry, and far-infrared space
interferometry. He is NASA Center Study Scientist for the Far-IR Surveyor, Mission Scientist
for the Wide-field Infrared Survey Explorer (WISE), and earlier served as Deputy Project
Scientist for the Cosmic Background Explorer (COBE) under Project Scientist and mentor Dr.
John Mather. He is Principal Investigator for “Wide-field Imaging Interferometry,” a Co-
Investigator on the “Balloon Experimental Twin Telescope for Infrared Interferometry
(BETTII),” and member of a three-person External Advisory Panel for the “Far Infrared Space
Interferometer Critical Assessment (FISICA),” a European Commission FP7 research program.
In 2004-05, he served as Principal Investigator and science team lead for the Space Infrared
Interferometric Telescope (SPIRIT) mission concept study. He has authored about 300
publications.

\vspace{2ex}\noindent\textbf{David J. Wilner} is a Senior Astrophysicist at the Smithsonian Astrophysical Observatory in
the Radio and Geoastronomy Division at the Center for Astrophysics, Harvard \&
Smithsonian. His main research interests are circumstellar disks and the formation of planets, and the development of aperture synthesis techniques. Much of his science program makes use
of radio, millimeter, and submillimeter interferometers, including the Submillimeter Array,
ALMA, and the VLA. He received an A.B. in Physics from Princeton University and a Ph.D. in
Astronomy from the University of California. He frequently lectures on imaging and
deconvolution in radio astronomy. He has authored about 450 publications.

\vspace{2ex}\noindent\textbf{Peter Roelfsema} is a senior scientist/project manager at SRON Netherlands Institute for Space
Research. He has been involved in several satellite projects, currently as PM for the Dutch
Athena/X-IFU contribution, and before that as PI for SPICA’s SAFARI Far-IR spectrometer and
as lead of the international SPICA collaboration. He was PI and ad-interim PI for Herschel/HIFI,
and in the early Herschel development phase, he was one of the lead system engineers
developing the Herschel ground segment concept and operational systems. Before Herschel he
led the ISO/SWS operations team in Villafranca/Spain and the SWS analysis software
development team. He started his scientific career as a radio astronomer, utilizing the WSRT, VLA and ATNF to study radio recombination lines of galactic HII regions and nearby active
galaxies. With ISO and Herschel he did (Far)IR spectroscopic work on galactic HII regions,
studying e.g. PAH properties and metal abundance variations in our galaxy. He has published
over 150 papers in astronomical journals conference proceedings and supervised a number of
PhD students.

\vspace{2ex}\noindent\textbf{Floris van der Tak} is a Senior Scientist in the Astrophysics program of the Netherlands Institute
for Space Research (SRON), where his research interests include astrochemistry, the habitability
of exoplanets, the physics of the interstellar medium, star formation, molecular spectroscopy and
radiative transfer. He received a PhD from Leiden University in 2000. He was the Project
Scientist for the SPICA/SAFARI instrument. He has authored about 216 publications.

\vspace{2ex}\noindent\textbf{Erick Young} is a Senior Science Advisor at Universities Space Research Association. He is a
widely recognized authority on infrared astronomy and the former Science Mission Operations
Director for SOFIA. He specializes in designing science instruments and has participated in
many NASA's space infrared astronomy missions. He was responsible for developing the far-
infrared detector arrays on the Spitzer Space Telescope’s Multiband Imaging Photometer for
Spitzer. As SOFIA Science Mission Operations Director, he manages the airborne observatory's
equipment, instruments, support facilities, and infrastructure. He was also responsible for the
overall scientific productivity of the facility, including the Guest Investigator program. He has
about 385 publications.

\vspace{2ex}\noindent\textbf{Christopher K. Walker} is a Professor of Astronomy, Optical Sciences, Electrical \& Computer Engineering, Aerospace \& Mechanical Engineering, and Applied Mathematics at the University of Arizona (UofA). He received his M.S.E.E. from Clemson University (1980), M.S.E.E. from Ohio State University (1981), and Ph.D. in Astronomy from the University of Arizona (1988). He has worked at TRW Aerospace and the Jet Propulsion Laboratory, was a Millikan Fellow in Physics at Caltech, and has been a faculty member at the UofA since 1991, where he has worked to advance the field of terahertz astronomy.  He has supervised sixteen Ph.D. students, led numerous NASA and NSF projects, authored/coauthored 130+ papers, and published two textbooks: "Terahertz Astronomy" and "Investigating Life in the Universe".

\listoffigures

\end{spacing}
\end{document}